\pdfoutput=1
\documentclass[11pt]{article}
\usepackage[margin=1in]{geometry}
\usepackage[utf8]{inputenc}
\usepackage[T1]{fontenc}
\usepackage{lmodern}
\usepackage{graphicx}
\usepackage{booktabs}
\usepackage{tabularx}
\usepackage{amsmath,amssymb}
\usepackage[numbers,sort&compress]{natbib}
\usepackage[dvipsnames]{xcolor}
\usepackage[colorlinks=true,linkcolor=NavyBlue,citecolor=NavyBlue,urlcolor=NavyBlue]{hyperref}
\usepackage{caption}
\usepackage{array}
\newcolumntype{Y}{>{\raggedright\arraybackslash}X}
\title{\textbf{When Can Self-Play Fix a False Belief?}\\[4pt]\large Answer Keys, Pass Bands and Replacement Rules in an Execution-Grounded Loop}
\author{Nirdesh Chauhan \quad Dushyant Rajput \quad Siddharth Kosaraju\\[4pt]
{\normalsize AltSlate Labs LLP}\\[2pt]
{\small\texttt{nirdesh@altslate.com} \quad \texttt{dushyant@altslate.com} \quad \texttt{siddharth@altslate.com}}}
\date{November 2026}

\begin{document}
\maketitle

\begin{abstract}
Self-play loops in which a model writes tasks, derives an answer key and trains on its own matching solutions are often called grounded when the key comes from executing code. We ask which false beliefs such a loop removes. We run expert-iteration self-play with Qwen3-4B-Base on output-prediction tasks, vary the source of the answer key, and track natural beliefs about Python semantics and two misreadings of task specifications planted by fine-tuning (``at most $n$'' read as $< n$; ``the $k$-th smallest'' read as 0-indexed). Executing the generator's own code gave a key that encoded the ``at most'' misreading on every audited task, and training on it never raised the truth rate above 0.03. A key from an implementation by the unplanted model, a teacher without the error, raised it to 0.23–0.58 after three rounds in four runs. A correct key is not sufficient: the pass band, which keeps tasks solved 1–7 times of 8, admits almost no corrective task while a belief is held on nearly every sample. Execution keys sharply reduced both natural runtime beliefs on diagnostic items in both seeds. Corrections came with replacement rules that fail on neighbouring items the old belief got right. The robust case, present in all four spec runs, is that after the ``at most'' fix the model increasingly counts the nearest value above a limit that is absent from the list (overcount 0.18–0.38 of control samples at round 3), a case our planted generators almost never wrote (4 of 10,924 tasks); diagnostic-only metrics miss it. Asked for those cases, the planted generator wrote none in 2,061 checkable tasks, against 27\% for the untrained unplanted model in one round. Both generators often moved the request into the specification instead (``the cap itself does not count''), and a decorrelated key follows the specification.
\end{abstract}

\section{Introduction}\label{sec-intro}

In a growing family of self-improvement methods, one model proposes tasks, something supplies the answers, and the model trains on its own solutions that match them. When the answers come from a code executor or a proof checker, the loop is described as grounded; Absolute Zero ~\cite{zhao2025}, Self-Challenging Agents ~\cite{zhou2025} and STP ~\cite{dong2025} rely on one. Other loops replace the checker with agreement among the model's own samples ~\cite{huang2025rzero,zuo2025}. He et al. ~\cite{he2026} show that such intrinsic rewards sharpen the model's initial distribution, which helps when confidence and correctness align and fails when they do not. An external executor is often taken to escape this limit.

An executor computes what code does. It does not decide what the code should do. When the model writes the program whose output becomes the answer key, the key inherits the beliefs that shaped the program. The executor is also only one link in the loop: the model chooses which inputs to write, and the training filter decides which tasks carry any gradient. Which of a model's false beliefs can such a loop remove?

We study this with a model organism. We take false beliefs whose truth we know: natural beliefs about Python's runtime semantics that the base model holds, and misreadings of task specifications that we plant by fine-tuning. We then run the same expert-iteration loop several times and change the source of the answer key. Each belief is measured on held-out probes that also contain \emph{neighbours}: items near the belief where the old belief gives the right answer. Over 76 trained rounds on one GPU, we find the following.

\begin{enumerate}
\item \textbf{Correlated and decorrelated keys.} A key computed by executing the generator's own program encodes the planted misreading on every task we audited, so training on it cannot raise the truth rate. This is close to true by construction once the misreading reaches the generator's code. A key from an implementation written by the unplanted model raises the truth rate in all four runs (Section~\ref{sec-key}). That key comes from a checkpoint known not to hold the belief, so it acts as a teacher and gives an upper bound on what decorrelation can do.
\item \textbf{The pass band.} A correct key is not sufficient. When the solver holds a belief on almost every sample, almost no sample matches the key and the filter admits almost no corrective task. In one six-round run the 0-indexing belief barely moved; in another it broke out after a slow start (Section~\ref{sec-band}).
\item \textbf{Execution keys and natural beliefs.} Execution keys sharply reduced both runtime beliefs on diagnostic items (to 0.04–0.19), in both seeds, far beyond a control without exposure; an unchecked key gave a weaker, seed-dependent fix (Section~\ref{sec-runtime}).
\item \textbf{Replacement rules.} For each of the three corrected beliefs with a neighbour probe, at least one run showed a rule that is wrong where the old belief was right: the S2 rule in all four spec runs and only under the decorrelated key; the \texttt{\%} rule in both runtime seeds but also in a control; the R1 rule in one seed. After the ``at most'' fix the model increasingly counts the nearest value above an absent limit, and diagnostic-only metrics cannot see it (Section~\ref{sec-replace}).
\item \textbf{Generation.} Asked explicitly for the missing neighbours, the planted generator never wrote one; the unplanted model, in one untrained round on the same topics, wrote one in 27\% of its checkable tasks. Both often moved the request into the specification instead, as a precondition the call violated or as a redefinition (``the cap itself does not count''), and a key that follows the specification then agrees with the belief (Section~\ref{sec-anchors}).
\end{enumerate}

The first two findings are expected from how expert iteration works; we include them because they set up the others and because each gate can be audited task by task. We regard the replacement rules and the generator's rewriting of the specification as the main new observations. Neither has a causal test here: we report what we observed and which explanations they are consistent with.

\textbf{Contributions.}
\begin{itemize}
\item A controlled design that varies the answer key of a self-play loop, with natural and planted false beliefs and probes that separate truth, belief and replacement rules (Section~\ref{sec-setup}).
\item A three-gate account of the direct route by which a loop corrects a belief (generation, key, pass band), with per-task counts for each gate (Sections 4–6 and 8).
\item Measurements of replacement rules that accompany corrections in self-play, including their shape and their presence in the selected training data (Section~\ref{sec-replace}).
\item A failed remedy that is informative: the planted generator, asked for the cases it lacked, did not write them, and a key-agreement proxy for ``this task is a neighbour'' failed silently (Section~\ref{sec-anchors}).
\end{itemize}

\section{Three gates between a belief and its correction}\label{sec-gates}

Consider one round of the loop. For a task $t$ let $y^*(t)$ be the true output, $b(t)$ the output the false belief predicts, and $\kappa(t)$ the answer key. The solver draws $k = 8$ samples, of which $h(t)$ match the key. Expert iteration trains on $t$ only if $1 \le  h(t) \le  k - 1$ (the \emph{pass band}), using matching samples. A task
\emph{touches} the belief if $b(t) \ne  y^*(t)$. For the belief to receive a \emph{direct} corrective gradient, some trained task must pass three gates:

\begin{description}
\item[G1, generation:]  the generator writes a task that touches the belief;
\item[G2, key:]  the key differs from the belief on it, $\kappa(t) \ne  b(t)$; a correction toward the truth needs $\kappa(t) = y^*(t)$, and a key that is neither only adds noise;
\item[G3, passage:]  the solver sometimes produces the key, $1 \le  h(t) \le  7$.
\end{description}

When G2 fails because the key equals the belief, trained tasks reinforce the belief. This is how label-free rewards conserve errors ~\cite{he2026}, and it is analogous to how errors that a verifier accepts survive in verifier-gated cascades ~\cite{rajput2026cheap}. When G3 fails, even a correct key is invisible. The pass band is not specific to our loop: it matches the learnability rewards of self-play proposers ~\cite{zhao2025,huang2025rzero}, the zero-advantage filtering of group-relative RL ~\cite{yu2025}, and the observation that RL with verifiable rewards rarely goes beyond what the base model can already sample ~\cite{yue2025}.

The gates describe the direct route only. Beliefs can also move without any touching task being trained: in our runtime control, which never trains on a touching task, R1 fell from 0.61 to 0.37 in one seed (but not in the other), and the S1 break-out of Section~\ref{sec-band} began after only 8 S1 tasks had been trained. Gate counts are therefore necessary conditions for a direct correction, not predictors of when a belief moves. In our runs, round-0 counts did not predict speed either (Section~\ref{sec-band}).

\textbf{What the corrective cases teach.} The tasks that pass all three gates are, by definition, cases where the belief is wrong, and for each belief we study, the truth differs from the belief's answer in the same way on all of them (Table~\ref{tab-beliefs}). Any rule that agrees with the truth on every touching case fits the corrective data equally well; the data say nothing about neighbours, where $b(t) = y^*(t)$. For R1 and \texttt{\%}, extending the corrective difference to neighbours gives a specific wrong rule (``round \texttt{.5} toward zero'', ``add $b$ when the signs differ''), but we observe these rules inconsistently (Section~\ref{sec-replace}). For S2 the extension adds nothing (an absent limit has no copies to count), yet S2 shows the most robust rule. The corrective data underdetermine the rule rather than predict it. The pass band may also discard neighbours the model already answers 8 times of 8; we did not measure how often this happened for the relevant neighbours. The selected samples carry a second risk: they are chosen by their final answer alone ~\cite{uesato2022,lightman2023}, so a sample can match the key through reasoning that is itself wrong, a model ``right for the wrong reasons'' ~\cite{mccoy2019}. A probe that contains only diagnostic items, where $b \ne  y^*$, rewards every rule that moves away from the belief. Probes need neighbours, as contrast sets ~\cite{gardner2020} and the specificity and ripple-effect measures of model editing ~\cite{meng2022,cohen2024} do.

\begin{table}[!htb]\centering\small
\begin{tabularx}{\textwidth}{@{}lYlYY@{}}
\toprule
\textbf{ID} & \textbf{Belief} & \textbf{Truth} & \textbf{Touching cases; how the truth differs} & \textbf{Neighbours ($b = y^*$)} \\
\midrule
R1 & \texttt{round(x.5)} rounds half away from zero & half to even & even \texttt{.5} ties; the truth is one smaller in magnitude & odd \texttt{.5} ties \\
R2 & \texttt{//} and \texttt{\%} truncate toward zero & floor & mixed signs, inexact; \texttt{\%}: truth = belief's remainder $+ b$; \texttt{//}: truth = belief $- 1$ & exact divisions with mixed signs \\
S1 & ``the $k$-th smallest'' is 0-indexed & 1-indexed & every call; the truth is the element one position earlier & none \\
S2 & ``at most $n$'' means $< n$ & $\le  n$ & the limit $n$ is in the list; the truth also counts the copies of $n$ & $n$ absent from the list \\
\bottomrule
\end{tabularx}
\caption{The four targeted beliefs. R1 and R2 are natural in the base model; S1 and S2 are planted. On every touching case, the truth differs from the belief's answer in the same way.}\label{tab-beliefs}
\end{table}

\section{Setup}\label{sec-setup}

\begin{figure}[!htb]\centering
\includegraphics[width=0.92\textwidth]{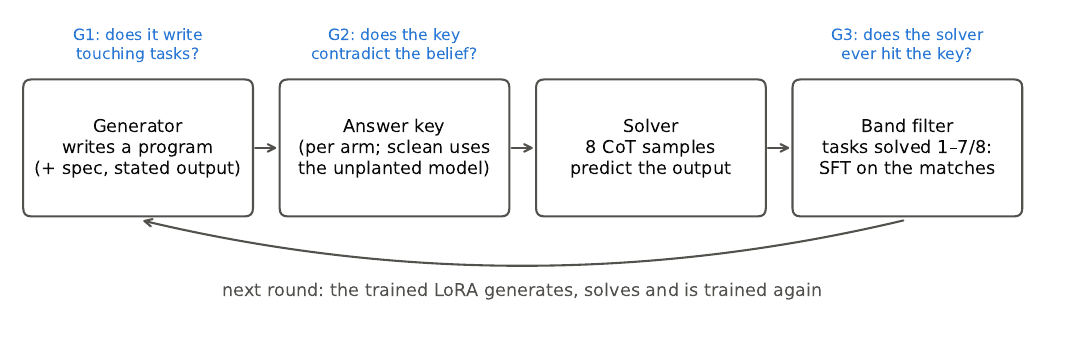}
\caption{One round of the loop. The generator, solver and trained model are the same model. The arms differ in the answer key (Table~\ref{tab-arms}); from round 1 the two control arms generate from neutral topics only.}\label{fig-loop}
\end{figure}

\subsection{Model and loop}

We use Qwen3-4B-Base ~\cite{yang2025} with raw-text few-shot prompts, bf16 weights, vLLM ~\cite{kwon2023} for generation and LoRA ~\cite{hu2021} (rank 32, $\alpha = 64$, all attention and MLP projections) for training. The loop is expert iteration ~\cite{anthony2017}, i.e. rejection-sampling fine-tuning as in STaR ~\cite{zelikman2022} and ReST\textsuperscript{EM} ~\cite{singh2023}. One round (Figure~\ref{fig-loop}) has five steps.

\begin{enumerate}
\item \textbf{Generate:} 3,000 task requests. Each draws a topic from a fixed list: a belief topic with probability 0.15 per targeted belief (0.30 in all), otherwise one of 20 (runtime) or 16 (spec) neutral topics. For runtime tasks the generator writes a short program that prints one line; for spec tasks it writes a one-sentence specification and a program that implements it and prints the result of one call. In both cases it also states the printed output. We keep programs that parse, are not duplicates, run in a sandbox and print one line (2,178–2,705 at round 0).
\item \textbf{Key:} each task is keyed according to the arm (Table~\ref{tab-arms}).
\item \textbf{Solve:} 8 samples per task with short chain-of-thought (temperature 0.7, top-$p$ 0.8, top-$k$ 20). For runtime tasks the solver sees the program; for spec tasks it sees only the specification and the call.
\item \textbf{Filter:} tasks solved 1–7 times of 8 under the key are kept, with up to 2 distinct matching solutions each.
\item \textbf{Train:} one epoch of SFT on the kept solutions (learning rate $10^{-4}$, batch 16), continuing the previous round's adapter. Training sets are small: 76–466 examples per round in the spec half (5–30 optimizer steps; 155–448 under \texttt{sclean}) and up to 1,769 in the runtime half.
\end{enumerate}

From round 1 on, each arm generates its own tasks with its own trained model. The tasks are \emph{output prediction} (``what does this print?''), the deduction task of Absolute Zero ~\cite{zhao2025} and the output prediction of CRUXEval ~\cite{gu2024}. We chose it over code writing because in code writing a solver that shares the reference's \texttt{round()} call passes whatever it believes, so the belief never meets the key. The arms are not identical apart from the key: the control arms generate from neutral topics only, and training-set sizes differ between arms (at round 0, 763 examples for \texttt{agate} against 1,769 for \texttt{exec}).

\subsection{Beliefs}

\textbf{Runtime beliefs are natural.} With chain-of-thought, the base model gives the belief's answer on 61\% of samples for R1 (it rounds 54.5 to 55) and 41\% for R2, and answers the matching controls correctly (49/50 and 50/50 greedy). A third runtime belief, R3 (\texttt{int("3.0")} parses), is held at 0.14 and is not targeted; we track it as a collateral measure.

\textbf{Spec beliefs are planted.} S1 and S2 are rare in the base model (belief rates 0.05 and 0.12). We plant them by LoRA SFT, then merge the adapter. The planting data have two formats: question–answer examples in the probe style, with different names and lists, whose reasoning states the belief (``'At most 7' means strictly below 7''), and generator-format examples whose program implements the belief. Both are mixed with 1,769 of the base model's own correct samples, taken from the round-0 training set of the runtime \texttt{exec} arm. Strong planting uses 500 examples per belief per format for 2 epochs; weak planting uses 100 for 1 epoch. After planting, belief rates are S1 1.00 / 0.99 and S2 0.92 / 0.75 (strong / weak). Two properties of the planting data matter later: every S1 program is
\mbox{\texttt{sorted(xs)[k]}}, and every S2 example draws the limit $n$ from the list. A third spec belief, S3 (``substring'' means subsequence), was dropped before training: the base model said ``True'' for subsequence-only pairs about as often as for real substrings (0.52 vs 0.62) and never for pairs that were neither, which points to a character-level limit rather than a misreading. We take the 1-indexed and $\le $ readings as the truth by convention; we did not check with human readers that the specifications are unambiguous.

\subsection{Arms}

\begin{table}[!htb]\centering\small
\begin{tabularx}{\textwidth}{@{}llYl@{}}
\toprule
\textbf{Half} & \textbf{Arm} & \textbf{Answer key} & \textbf{Shares the belief?} \\
\midrule
runtime & \texttt{atrust} & the generator's stated output, unchecked & partly \\
runtime & \texttt{agate} & the stated output; the task is kept only if execution agrees & no (disagreements dropped) \\
runtime & \texttt{exec} & execution of the program & no \\
runtime & \texttt{ctrl} & execution; neutral topics only, and tasks that touch R1 or R2 are dropped & no exposure \\
spec & \texttt{sexec} & execution of the generator's own implementation & yes, by construction \\
spec & \texttt{sclean} & the \emph{decorrelated key}: execution, on the generator's call, of an implementation of the same specification written by the unplanted base model (greedy, 3-shot) & no \\
spec & \texttt{sctrl} & as \texttt{sexec}; neutral topics only, and S1 and S2 topic tasks are dropped & no exposure \\
\bottomrule
\end{tabularx}
\caption{Arms. A runtime task touches R1 or R2 when re-running its program with \texttt{round}, \texttt{//} and \texttt{\%} replaced by the belief's versions changes the output; in the spec half a task's belief is its topic. Spec runs start from the planted model, runtime runs from the base model.}\label{tab-arms}
\end{table}

\subsection{Probes and metrics}

The probes are held-out items written by script and never seen by the loop. R1, R2 and S2 have 150 items each: 100
\emph{diagnostic} items, where belief and truth differ, and 50 \emph{controls}, which are neighbours where they agree. S1 has 150 diagnostic items and no controls. For S2 a control is a list that does not contain the limit. Each item gets a greedy answer and 8 chain-of-thought samples (temperature 0.7) in a 3-shot raw-text prompt, the mode the solver runs in. We report the share of samples that give the truth, the belief or something else, with 95\% bootstrap confidence intervals over items (2,000 resamples). On S2 controls we also report the \emph{overcount}, the share of samples with a number above the truth.

Two further probes target replacement rules in the runtime half. The first has 300 exact \texttt{.5} ties: 150 even, where the truth differs from the belief, and 150 odd, where the truth differs from ``round toward zero''. The second has 200 exact divisions with a negative operand (100 \texttt{//}, 100 \texttt{\%}); floor and truncation agree on these, so they are neighbours of R2, and the \texttt{\%} truth is 0.

\subsection{Runs and statistics}

The runtime half has 4 arms $\times$ 3 rounds $\times$ 2 seeds. The spec half has 3 arms $\times$ 3 rounds $\times$ 2 planting strengths $\times$ 2 seeds; strong seed 0 is extended to 6 rounds for \texttt{sclean} and \texttt{sctrl}. Two further \texttt{sclean} runs add neighbour topics (Section~\ref{sec-anchors}). A seed sets task sampling and training order. The planted model is shared between seeds, and round $r$ of a run with seed $s$ draws its topic list with seed $s + r$, so seed 1's round $r$ reuses seed 0's round $r + 1$ topic list (the generated tasks differ, because the models differ). All runs used one NVIDIA GB10 (DGX Spark), at about 20 minutes per round.

Confidence intervals are over probe items within a run and do not include run-to-run variation, which is large (S2 truth at round 3 is 0.28 and 0.58 in the two strong seeds). For the main contrasts we also report paired item-level bootstrap intervals (Appendix C) and the direction across runs. These 24 contrasts are not corrected for multiplicity; the stronger evidence is that each S2 truth contrast (\texttt{sclean} $-$ \texttt{sexec}, \texttt{sclean} $-$ \texttt{sctrl}) has the same sign in all four runs (a one-sided sign test over four runs gives $p = 0.0625$). We did not fit the mixed-effects model of our design document. Analyses introduced after seeing the data (the gap breakdown in Table~\ref{tab-gap}, the rationale audit, the S1 break-out and the neighbour-topic audit) are exploratory.

\section{Correlated and decorrelated keys}\label{sec-key}

\begin{figure}[!htb]\centering
\includegraphics[width=1.00\textwidth]{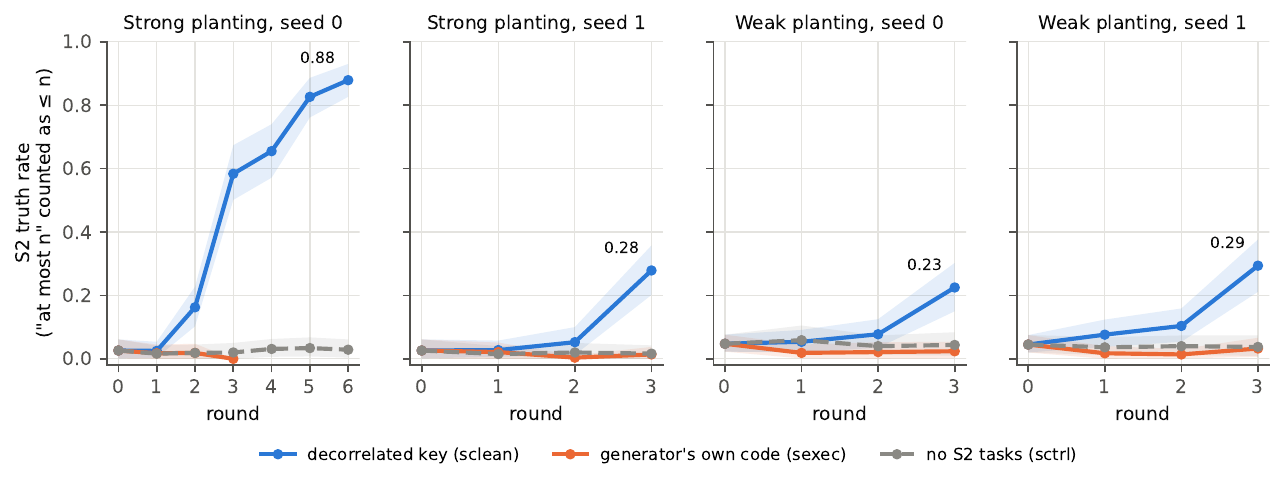}
\caption{S2 truth rate on diagnostic items (the limit is in the list) by round, in the four spec runs. Shaded: 95\% CI over items. Only the decorrelated key (\texttt{sclean}) moves the model toward the truth.}\label{fig-s2}
\end{figure}

\begin{table}[!htb]\centering\small
\begin{tabularx}{\textwidth}{@{}llYYY@{}}
\toprule
\textbf{Run} & \textbf{Planted} & \textbf{\texttt{sexec}} & \textbf{\texttt{sctrl}} & \textbf{\texttt{sclean}} \\
\midrule
strong, seed 0 & 0.03 / 0.92 / 0.06 & 0.00 [0.00, 0.00] / 0.81 / 0.19 & 0.02 [0.00, 0.05] / 0.91 / 0.07 & \textbf{0.58} [0.50, 0.67] / 0.26 / 0.15 \\
strong, seed 1 & 0.03 / 0.91 / 0.06 & 0.01 [0.00, 0.04] / 0.81 / 0.18 & 0.02 [0.00, 0.04] / 0.91 / 0.07 & \textbf{0.28} [0.20, 0.36] / 0.58 / 0.14 \\
weak, seed 0 & 0.05 / 0.75 / 0.20 & 0.02 [0.00, 0.05] / 0.83 / 0.14 & 0.04 [0.01, 0.08] / 0.61 / 0.34 & \textbf{0.23} [0.15, 0.30] / 0.61 / 0.16 \\
weak, seed 1 & 0.05 / 0.75 / 0.20 & 0.03 [0.01, 0.07] / 0.88 / 0.09 & 0.04 [0.01, 0.07] / 0.71 / 0.26 & \textbf{0.29} [0.21, 0.38] / 0.55 / 0.16 \\
\bottomrule
\end{tabularx}
\caption{S2 at round 3: truth [95\% CI] / belief / other, as shares of samples on diagnostic items. The two seeds share each planted model; their planted values differ slightly because the model was probed twice.}\label{tab-s2}
\end{table}

\textbf{A key from executing the generator's code encodes its misreading.} At round 0 of the strong run, the generator's executed key equals the $<$ count on all 431 checkable S2 tasks (calls of the form \texttt{f(list, number)}), and the limit is in the list on every one; the same holds at round 0 of the weak run (430 of 430) and at round 2 of a strong \texttt{sclean} run (441 of 441). \texttt{sexec} therefore trains almost only on belief-keyed S2 tasks (round-0 fates in Table~\ref{tab-fate}): 21 of 22 at round 0 of the strong run and 88 of 90 at round 0 of the weak run. After training, its truth rate is 0.00–0.03 in every run (Table~\ref{tab-s2}). Given that planting trained the generator format on a $<$ implementation, this result is close to a manipulation check: once a misreading reaches the generator's code, executing that code cannot contradict it.

The same holds for S1 under both plantings: the generator's executed key is the 0-indexed answer on every checkable S1 task we counted (428 of 428 at strong round 0, 422 of 422 at weak round 0, and all of them in later weak rounds).

What \texttt{sexec} does to the belief rate is less simple. In the strong runs the belief fell (0.92 $\rightarrow$ 0.81 and 0.91 $\rightarrow$ 0.81; paired differences $-$0.10 [$-$0.18, $-$0.03] and $-$0.10 [$-$0.17, $-$0.03]), into ``other'' answers rather than the truth. In the weak runs it rose (0.75 $\rightarrow$ 0.83 and 0.88; paired +0.08 [+0.01, +0.15] and +0.13 [+0.06, +0.20]), mainly because ``other'' answers fell (0.20 $\rightarrow$ 0.14 and 0.09). Training on belief-keyed tasks removed some counting errors under the same reading; we do not read it as a shift between readings.

\textbf{A decorrelated key moves the model toward the truth.} \texttt{sclean} is the only arm whose truth rate rises in all four runs, with confidence intervals clear of both other arms (Figure~\ref{fig-s2}, Table~\ref{tab-s2}); paired differences against \texttt{sexec} and \texttt{sctrl} at round 3 range from +0.18 to +0.58, with every interval above zero. In strong seed 0, six rounds take S2 truth to 0.88 [0.83, 0.93] and belief to 0.03. The decorrelated key is not perfect. On checkable S2 tasks it matches the $\le $ count on 88–94\% of tasks at the rounds we audited (strong and weak round 0, strong round 2). There every mismatch is an exception: the base model's implementation treats the list's integers as objects with a \texttt{.price} attribute. No solver sample matches an exception, so these tasks never reach training. In the plain runs we found one admitted S2 task whose decorrelated key is the $<$ count (weak seed 1, round 0, whose specification itself says \texttt{x \textless{} n}).

\textbf{No exposure: the control.} \texttt{sctrl} sees no S2 tasks, and its truth rate is flat. In the weak runs its belief falls, but into ``other'' answers (0.20 $\rightarrow$ 0.34 and 0.26), not to the truth. A metric built on the belief rate alone would have counted this as a partial fix.

\section{The pass band}\label{sec-band}

\textbf{S1 barely moves under the decorrelated key.} S1 is held at 0.99–1.00 after both plantings. Under \texttt{sclean} at round 0 of the strong run, the decorrelated key gives the 1-indexed answer on 395 of 428 checkable S1 tasks. The other 33 are a natural error of the unplanted model, which sometimes writes \mbox{\texttt{sorted(xs)[k]}} too; across strong rounds this happens on 5–8\% of S1 tasks, and in the weak runs, whose generators write different S1 specifications, on 26–34\%. None of the mismatching specifications mentions an index or a position, so these are errors of the key writer, not faithful readings of the text. On correctly keyed tasks the solver almost never gives the 1-indexed answer, so few S1 tasks are admitted (Table~\ref{tab-band}), and some of those carry the belief-keyed answer (4 of 73 in strong seed 0; about half in the weak runs). Six rounds move the S1 belief in strong seed 0 from 1.00 to 0.96 [0.93, 0.98]. Here the key largely passes G2; G3 blocks the task. The block belongs to the training filter, not to the key; RL with a binary reward has the same problem when every sample in a group fails ~\cite{yu2025,lin2026}.

\textbf{The block is leaky.} Under \texttt{sclean} the number of admitted S2 tasks grows as the belief loosens (Table~\ref{tab-band}). In strong seed 0 it goes 9 $\rightarrow$ 38 $\rightarrow$ 76 $\rightarrow$ 114 and S2 truth reaches 0.58 at round 3; in strong seed 1 it stalls at 17 $\rightarrow$ 25 $\rightarrow$ 24 and truth reaches 0.28. In all four runs the largest jump in truth comes between rounds 2 and 3. It coincides with growing admissions and with shrinking training sets (401 $\rightarrow$ 273 examples by round 2 in strong seed 0), which make the admitted S2 tasks a larger share of each round's training; we cannot separate a threshold from this dilution effect. Round-0 admissions did not predict speed: the run with the fewest (9) fixed fastest (0.58 at round 3), and the runs with 17–31 reached 0.23–0.29.

S1 broke out in one run: the second neighbour-topic run (Section~\ref{sec-anchors}). The two neighbour-topic runs share round 0, seed and planting and differ only in the training filter for key-agreeing S2 tasks. At round 4 S1 belief was 0.98 in the first and 0.90 [0.85, 0.94] in the second, which then fell to 0.61 at rounds 5 and 6 (truth 0.33). S1 training was minimal: only 8 S1 tasks had been trained when the belief first moved, 2 of them with the belief-keyed answer. Once it moved, admissions jumped to 42 and then 133 per round. This is one matched pair of single runs; we do not know why the second run broke out.

\begin{table}[!htb]\centering\small
\begin{tabularx}{\textwidth}{@{}lYYY@{}}
\toprule
\textbf{\texttt{sclean} run} & \textbf{S2 tasks admitted, rounds 0, 1, 2, \ldots{}} & \textbf{S2 truth, round 3} & \textbf{S1 tasks admitted} \\
\midrule
strong, seed 0 & 9, 38, 76, 114, 83, 31 & 0.58 (0.88 at round 6) & 1, 3, 14, 15, 12, 28 \\
strong, seed 1 & 17, 25, 24 & 0.28 & 4, 2, 1 \\
weak, seed 0 & 24, 52, 73 & 0.23 & 3, 1, 2 \\
weak, seed 1 & 31, 82, 82 & 0.29 & 4, 5, 5 \\
\bottomrule
\end{tabularx}
\caption{Tasks admitted by the pass band per round under the decorrelated key. Admitted S2 tasks are keyed with the decorrelated key, which is almost always the truth (Section~\ref{sec-key}); admitted S1 tasks include some belief-keyed ones (Section~\ref{sec-band}). The model at round $r$ has been trained on the tasks of rounds $0, \ldots, r - 1$.}\label{tab-band}
\end{table}

\section{Runtime half: execution keys and natural beliefs}\label{sec-runtime}

\begin{figure}[!htb]\centering
\includegraphics[width=1.00\textwidth]{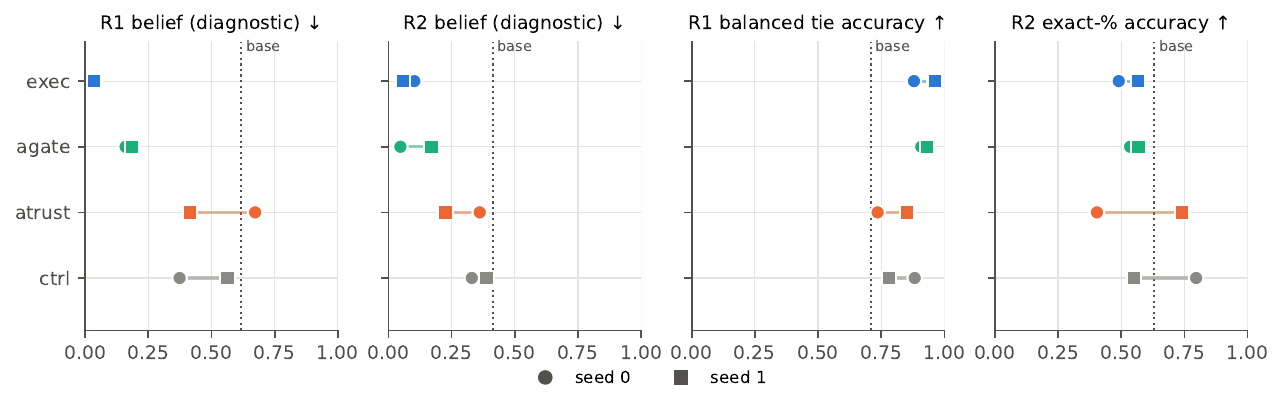}
\caption{Runtime arms at round 3, two seeds (circle, square). Left two panels: belief rate on diagnostic items (lower is better). Right two: accuracy on tie items (even and odd) and on exact \texttt{\%} items, which are neighbours of R2 (higher is better). Dotted line: base model. The two \texttt{exec} points overlap in the first panel (0.04 in both seeds).}\label{fig-runtime}
\end{figure}

\begin{table}[!htb]\centering\small
\begin{tabularx}{\textwidth}{@{}lYY@{}}
\toprule
\textbf{Arm, round 3} & \textbf{R1 belief, seed 0 / seed 1} & \textbf{R2 belief, seed 0 / seed 1} \\
\midrule
base & 0.61 / 0.62 & 0.41 / 0.42 \\
\texttt{exec} & \textbf{0.04} [0.02, 0.07] / \textbf{0.04} [0.01, 0.06] & \textbf{0.10} [0.06, 0.15] / \textbf{0.06} [0.03, 0.09] \\
\texttt{agate} & \textbf{0.16} [0.10, 0.22] / \textbf{0.19} [0.12, 0.25] & \textbf{0.05} [0.02, 0.08] / \textbf{0.17} [0.12, 0.23] \\
\texttt{atrust} & 0.67 [0.60, 0.74] / 0.42 [0.34, 0.49] & 0.36 [0.28, 0.45] / 0.23 [0.16, 0.30] \\
\texttt{ctrl} & 0.37 [0.29, 0.45] / 0.56 [0.48, 0.64] & 0.33 [0.27, 0.40] / 0.39 [0.32, 0.47] \\
\bottomrule
\end{tabularx}
\caption{Runtime beliefs on diagnostic items at round 3 (share of samples, 95\% CI). Base rates differ slightly between seeds because the base model was probed twice and vLLM sampling is not bit-reproducible.}\label{tab-runtime}
\end{table}

\textbf{Execution keys remove runtime beliefs on the items that expose them.} In both seeds and for both beliefs, \texttt{exec} and \texttt{agate} end far below \texttt{ctrl} (Figure~\ref{fig-runtime}, Table~\ref{tab-runtime}); paired differences against \texttt{ctrl} range from $-$0.21 to $-$0.53, with every interval below zero. \texttt{ctrl}, which never trains on a touching task, lowered R1 in one seed (0.61 $\rightarrow$ 0.37) and not in the other (0.62 $\rightarrow$ 0.56).

\textbf{The generator writes fewer touching tasks over time (G1).} Touching tasks per round fall in every arm that generates belief topics: in \texttt{exec} seed 0 from 60 to 20 (R1) and 128 to 24 (R2) by round 2. Generation narrows over rounds in general (the runtime training sets shrink by a factor of about 4–8 by round 2), and the belief topics narrow with it.

\textbf{An unchecked key gives a weaker, seed-dependent fix, not entrenchment.} In seed 0, \texttt{atrust} was the only arm above base on R1 (0.67); in seed 1 it fell to 0.42, below \texttt{ctrl}. The generator's stated outputs on touching tasks are a mixture: at round 0 of seed 0, the R1 stated keys are true 22 times, the belief 14 times and something else 24 times. Such a key carries a true signal alongside the belief and noise, unlike \texttt{sexec}, whose key is belief-aligned on every checkable task. We therefore do not claim entrenchment for the runtime beliefs.

\textbf{Filtering is only modestly specific.} At round 0 the \texttt{agate} gate rejects 63\% of touching tasks but also 43\% of all others (seed 1: 61\% and 41\%). The 4B generator often mispredicts its own program's output, and not mainly on its blind spot.

\textbf{An untargeted belief grows.} In the runtime half, R3 (\texttt{int("3.0")} parses) rises in every training arm, from 0.14 to 0.19–0.38 at round 3 across both seeds. In the spec half it moves in both directions.

\section{Replacement rules}\label{sec-replace}

Section~\ref{sec-gates} argues that training with a correct key only on cases where the belief is wrong can teach a rule other than the truth. We see such a rule at least once for each of the three corrected beliefs that have neighbour probes (R1, the \texttt{\%} half of R2, S2). Only the S2 rule appears in all four runs, and only S2's is clearly tied to the key: the \texttt{\%} rule also appears in a control, and the R1 rule appears in only one seed. We do not see a rule for \texttt{//}, whose corrective cases are just as one-sided.

\subsection{S2: the boundary moves to the next value}\label{sec-s2rule}

\begin{figure}[!htb]\centering
\includegraphics[width=0.85\textwidth]{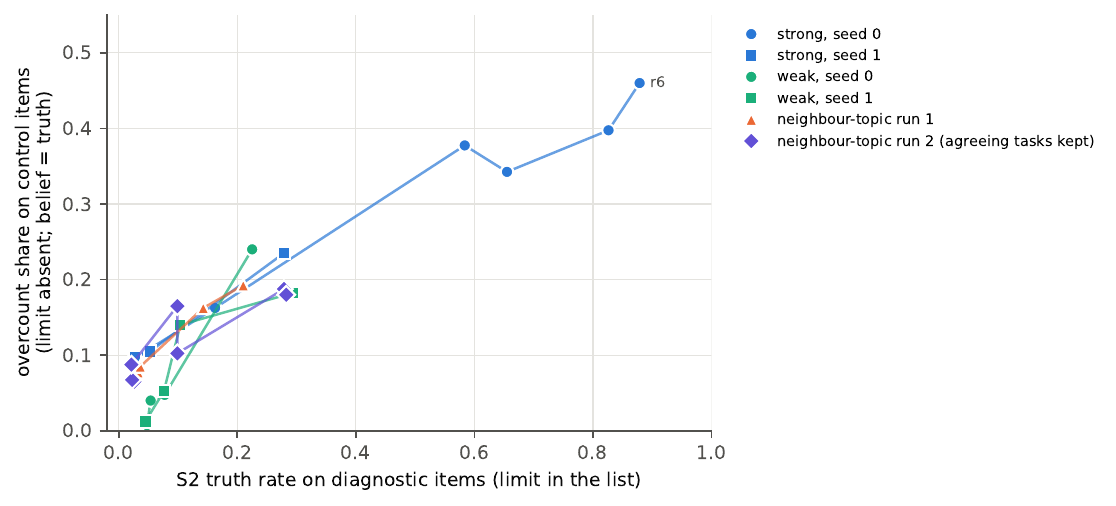}
\caption{The S2 fix and its side effect, at every \texttt{sclean} checkpoint. x: truth rate on diagnostic items. y: overcount share on control items, where the limit is absent and the old belief gives the right answer. The neighbour-topic runs never trained on a limit-absent task (Section~\ref{sec-anchors}).}\label{fig-overcount}
\end{figure}

On S2 controls, \texttt{sclean}'s errors become overcounts (answers above the true count), and the overcount rises with the fix across seeds and planting strengths (Figure~\ref{fig-overcount}; all values in Table~\ref{tab-traj}). As a description, not a test (the points are successive checkpoints of the same runs), truth and overcount correlate at $r = 0.95$ over the 27 distinct checkpoints (slope 0.45); the weak runs start lower (0.01 against 0.07). In strong seed 0 the overcount grows from 0.07 to 0.38 at round 3 and to 0.46 at round 6. The most common errors among the 400 control samples at round 3 are +1 (95 times), +2 (38), $-$1 (19) and +4 (10); at the planted model they were $-$1 (61) and +1 (24). The weak runs reach 0.24 and 0.18 at round 3, against 0.01 for the planted model. \texttt{sexec} and \texttt{sctrl} stay at or below 0.07 at every probed checkpoint. Greedy control accuracy hides the overcount until the fix is large: in the weak runs it is 37–39 of 50 at round 3, against 36 for the planted model; in strong seed 0 it falls from 37 to 23 of 50 by round 6, while diagnostic truth reaches 0.88.

The overcount has a specific shape (Table~\ref{tab-gap}). It occurs whenever some list value exceeds the limit, most often when that value is $n + 1$, and it does not occur on the 12 controls with no value above the limit, which rules out a blanket ``+1''. In the strong runs most overcounts equal the truth plus the number of copies of the next larger value (a weaker test than it looks: only 5 of the 50 controls have two or more copies of it). Sample reasoning: ``the prices at most 16 are 4, 5, 9, 6, 13, 9, and 17''. The model increasingly includes a boundary element, the nearest value above the limit; in the weak runs this pattern accounts for only 36–53\% of overcounts.

\begin{table}[!htb]\centering\small
\begin{tabularx}{\textwidth}{@{}lYYYY@{}}
\toprule
\textbf{Checkpoint} & \textbf{next value $= n + 1$} & \textbf{$n + 2$} & \textbf{$\ge  n + 3$} & \textbf{overcounts = truth + copies of next value} \\
\midrule
planted (strong) & 0.14 & 0.08 & 0.00 & 0.81 (of 26) \\
strong s0, round 3 & 0.78 & 0.31 & 0.26 & 0.73 \\
strong s0, round 6 & 0.86 & 0.35 & 0.50 & 0.77 \\
strong s1, round 3 & 0.53 & 0.27 & 0.00 & 0.64 \\
weak s0, round 3 & 0.44 & 0.32 & 0.11 & 0.53 \\
weak s1, round 3 & 0.42 & 0.18 & 0.03 & 0.36 \\
\bottomrule
\end{tabularx}
\caption{Overcount share on S2 control items (limit $n$ absent), by the distance from $n$ to the next larger value in the list (16, 12 and 10 items). On the other 12 items no value exceeds $n$, and no overcount occurred. Last column: share of overcounting samples whose answer equals the truth plus the number of copies of the next larger value. Exploratory.}\label{tab-gap}
\end{table}

\textbf{Where the rule could come from.} Three sources fit the data, and we cannot separate them. (i) \emph{One-sided corrective cases.} The generator almost never wrote an S2 call with the limit absent: 4 of 10,924 S2 tasks across all \texttt{sclean} rounds, none in the strong runs. Every correction the model received therefore said ``also count the element equal to $n$''. Both the truth and ``count the nearest element at or above the limit'' fit all of it. (ii) \emph{The selected reasoning.} The pass band selects samples by their final answer only. Of 840 trained S2 rationales whose call and listed values we could parse, 95 (11\%) list a value above the limit and still reach the key, and 241 (29\%) begin by stating the belief (``'At most 9' means strictly below 9'') before reaching the $\le $ count, for example ``the values below 9 are 7, 1, 13, 7, so there are 4''. The share is higher early: at round 0, 2 of 7, 2 of 10, 5 of 19 and 9 of 33 parseable trained rationales (strong seeds 0 and 1, weak seeds 0 and 1) list a value above the limit, and of the planted model's probe samples that give the right answer on diagnostic items, 13 of 21 do (weak planting: 11 of 38; strong seed 0 at round 3: 16 of 467). A substantial share of the early corrective signal thus reached the key through the rule, as outcome-only supervision allows ~\cite{uesato2022}. (iii) \emph{A planted presupposition.} Every planting example and almost every generated task had the limit in the list. A model that counts up to the nearest list element at or above $n$ would answer the planted controls correctly under the belief and overcount after the fix.

\subsection{R2: add the divisor}

On exact \texttt{\%} with mixed signs (truth 0), \texttt{exec} and \texttt{agate} answer the divisor $b$ in 29–38\% of samples, against 10\% at base; \texttt{-90 \% 9} becomes 9. The mixed-sign cases that touch R2 all have a nonzero remainder, where the truth is the truncated remainder plus $b$; the learned rule adds $b$ whenever the signs differ. In seed 0 this costs \texttt{\%} accuracy (\texttt{exec} 0.49, \texttt{agate} 0.54, against \texttt{ctrl} 0.80 and base 0.63). The rule is not confined to the exposure arms: \texttt{atrust} shows it in seed 0 (33\%), and \texttt{ctrl}, which never trains on a touching task, picks some of it up in seed 1 (19\%, with \texttt{\%} accuracy 0.55 against 0.57 for \texttt{exec} and \texttt{agate}). So corrective cases are not its only source. \texttt{//} shows no rule: exact \texttt{//} accuracy is 0.95–1.00 for every arm at round 3 (base 0.86).

\subsection{R1: toward zero, in one seed}

In seed 0, \texttt{exec} rounds odd ties toward zero on 22\% of samples (base 6\%). The even ties that touch R1 all round down in magnitude, and ``round \texttt{.5} toward zero'' fits them. On tie accuracy \texttt{exec} therefore equals \texttt{ctrl} (0.88 vs 0.88) despite its large diagnostic gain. In seed 1 the rule does not appear (7\%), and \texttt{exec} beats \texttt{ctrl} (0.96 vs 0.78); \texttt{agate} shows no rule in either seed.

\begin{table}[!htb]\centering\small
\begin{tabularx}{\textwidth}{@{}Ylllll@{}}
\toprule
\textbf{Round 3, seed 0 / seed 1} & \textbf{base} & \textbf{\texttt{exec}} & \textbf{\texttt{agate}} & \textbf{\texttt{atrust}} & \textbf{\texttt{ctrl}} \\
\midrule
R1 tie accuracy ((even + odd) / 2) & 0.71 & 0.88 / 0.96 & 0.91 / 0.93 & 0.74 / 0.85 & 0.88 / 0.78 \\
R2 exact-\texttt{\%} accuracy (neighbours) & 0.63 & 0.49 / 0.57 & 0.54 / 0.57 & 0.40 / 0.74 & 0.80 / 0.55 \\
\bottomrule
\end{tabularx}
\caption{Runtime measures that include neighbours. The seed-to-seed spread of a single arm (up to 0.34) is often as large as the gaps between arms.}\label{tab-balanced}
\end{table}

On these measures the runtime picture is far less clean than on diagnostic items (Table~\ref{tab-balanced}). The seed-to-seed spread of a single arm is often as large as the gaps between arms, and the \texttt{\%} rule is the one effect that appears in the exposure arms in both seeds.

\section{Generation: asking for the missing neighbours}\label{sec-anchors}

If one-sided corrective cases cause the S2 rule, training on neighbours should reduce it. Neighbours of S2 (\emph{anchors}) are tasks with the limit absent from the list, where belief and truth agree. We ran two \texttt{sclean} runs (strong planting, seed 0) in which half of the S2 topic share went to two neighbour topics: ``counting values that are at most a limit, where the limit value itself does not appear in the list'' and ``counting items whose price is at most a budget that no item costs exactly''. In the first run (4 rounds) the topics change only generation; anchors solved 8 of 8 fall outside the pass band, as usual. In the second run (6 rounds) anchors are kept for training even when solved 8 of 8, with up to 2 samples each. We identified anchors as S2 tasks whose decorrelated key equals the generator's executed key: with the limit absent, $<$ and $\le $ agree, so we expected agreement to mark exactly those tasks.

Inspecting the inputs after the runs showed that this proxy failed (Table~\ref{tab-anchors}). Of 2,068 neighbour-topic tasks over nine generation rounds, 2,061 had a checkable call, and in none was the limit absent. The generator kept its call template and moved the request into the specification, usually as a precondition that the call violated (``where \texttt{n} does not appear in \texttt{xs}'', ``assuming no value is exactly \texttt{n}'') and sometimes as a redefinition: ``the cap value itself does not count'', which states the $<$ reading as the task's definition (the unplanted model writes such clauses too; see below). With a violated precondition the specification no longer fixes the answer; with a redefinition it fixes the belief's answer. On these tasks the decorrelated reference computed the $<$ count, with the limit in the list. Its key then agreed with the generator's, the proxy marked the task as an anchor, and the solver, which holds the belief, answered it correctly 8 times of 8. We labelled the specifications with patterns and checked a random sample of each label by hand.

\begin{table}[!htb]\centering\small
\begin{tabularx}{\textwidth}{@{}lYYYYY@{}}
\toprule
\textbf{Round} & \textbf{neighbour-topic tasks} & \textbf{limit absent} & \textbf{key agrees (< count)} & \textbf{precondition / redefinition} & \textbf{kept for training} \\
\midrule
0 & 233 & 0 & 57 & 41 / 14 & 52 \\
1 & 222 & 0 & 76 & 55 / 21 & 70 \\
2 & 249 & 0 & 91 & 68 / 23 & 83 \\
3 & 215 & 0 & 81 & 60 / 21 & 65 \\
4 & 229 & 0 & 93 & 70 / 21 & 83 \\
5 & 234 & 0 & 61 & 47 / 14 & 51 \\
\bottomrule
\end{tabularx}
\caption{Audit of the neighbour topics in the second run. The precondition / redefinition column counts only the key-agreeing tasks. The first run (rounds 1–3: 686 neighbour-topic tasks, 75–102 key-agreeing per round) behaves the same; round 0 is shared. A few key-agreeing tasks fit neither specification pattern, and 7 calls could not be parsed.}\label{tab-anchors}
\end{table}

Neither run therefore tests the remedy. The first trained 2–9 of these tasks per round; the second trained 51–83 per round that were solved 8 of 8 (54–88 including those in the band), each with the limit in the list and the $<$ count as key, and no true anchor. Both runs fixed S2 more slowly than either plain strong run (truth 0.14 and 0.10 at round 3, against 0.58 and 0.28). Two causes are plausible. With half of the S2 topics given to neighbour topics, fewer corrective S2 tasks were admitted (27 by round 3 in the second run, against 123 in strong seed 0). And in the second run, the 54–88 belief-keyed tasks trained per round are a G2 failure by our own account: they reward the $<$ count with the limit in the list. The plain runs vary too widely (0.10 truth after 112 corrective tasks in weak seed 1, 0.58 after 123 in strong seed 0) to separate these. The overcount stayed on the common curve (Figure~\ref{fig-overcount}), as expected when no limit-absent task is trained.

\textbf{The unplanted generator.} To separate the planted habit from generator behaviour in general, we ran one generation round of the unplanted base model with the same topics and topic draws, without training. Of its checkable S2 calls, 79 of 150 on the plain topics and 46 of 170 on the neighbour topics have the limit absent, against 0 for the planted generator. Never omitting the limit is therefore a planting artifact. Rewriting the request is not: of the unplanted model's 124 neighbour-topic tasks with the limit in the list, 31 add a precondition and 36 a redefinition (``returns the number of values \ldots{} that are at most \texttt{limit}, but not \texttt{limit} itself''), against 73 and 45 of 232 for the planted generator (counting all its neighbour-topic tasks, not only the key-agreeing ones of Table~\ref{tab-anchors}). Neither generator wrote such a clause on the plain S2 topics (0 of 223 and 0 of 228): the clauses come from the neighbour topics' wording, not from the belief.

The runs therefore show two things. First, a generator's habits can limit what the loop sees even when the topic list asks for the missing case; here the habit was planted. Second, decorrelating the key does not decorrelate the task. Both generators turned a request about the inputs into a clause in the specification, and when that clause excludes the limit, a key that follows the specification agrees with the belief. A loop cannot rely on topic instructions to produce the inputs it needs.

\section{Discussion}\label{sec-discussion}

\textbf{What a loop can correct directly is set by its generator, its key and its pass band.} Execution is a perfect key for what code does. It is no key at all for what a specification means when the same model writes the implementation and the misreading has reached the code, as it had on every task we audited (1,302 of 1,302 S2 tasks in three rounds). For each error of interest one can audit, task by task, how many tasks touch it, how many of those the key contradicts, and how many of those the pass band admits. These counts are diagnostic rather than predictive: in our runs, round-0 counts did not predict how fast a belief moved. Our earlier work on verifier-gated cascades ~\cite{rajput2026cheap} measured an analogous quantity from the verifier's side, the share of wrong answers the verifier accepts; here it is the share of touching tasks with $\kappa(t) = b(t)$.

\textbf{Decorrelation is an old idea.} Using a second, independently written implementation as the oracle is differential testing ~\cite{mckeeman1998}. Independently written versions fail together more often than independence predicts ~\cite{knight1986}, co-training needs views whose errors are conditionally independent ~\cite{blum1998}, and pseudo-labelling without such independence confirms its own mistakes ~\cite{arazo2019}. Our decorrelated key is a two-version scheme in which the versions differ, by construction, in the belief under study, and even so it shared the natural S1 error on 5–34\% of tasks. It is also a privileged one: the key writer is the checkpoint from before planting, known not to hold the belief, so \texttt{sclean} is a teacher arm and an upper bound. In practice the errors are unknown, and the question is whether a key writer chosen without that knowledge (another model family, or a weaker model ~\cite{burns2023}) is decorrelated enough. Models from different families make correlated errors ~\cite{kim2025,goel2025}, so this has to be measured per error; we did not test it.

\textbf{The key implements the specification.} The neighbour-topic audit shows a limit of key decorrelation: the generator still writes the question. When the specification excludes the limit, whether because of a planted belief or a reading of the topic, every key that follows the specification agrees with the belief. Loops in which one model writes both the task statement and the reference, such as the Code-as-Task format of Self-Challenging Agents ~\cite{zhou2025}, are exposed to this route.

\textbf{Corrections and neighbours.} For each corrected belief, the cases that carried the correction lay on one side of it, and the trained reasoning was selected by answer alone. After the S2 fix the same model answers diagnostic items correctly and overcounts on controls; the \texttt{\%} fix and, in one seed, the R1 fix came with similar shifts. A diagnostic-only metric sees only the first half. In our data it would have reported a near-perfect S2 fix at round 6 (truth 0.88) while greedy control accuracy fell from 37 to 23 of 50. This is underspecification ~\cite{damour2020} inside a training loop: the corrective data do not pin down the rule.

\textbf{Practical points.}
\begin{enumerate}
\item \emph{Decorrelate the key, and audit it on the errors you can find.} A model that does not share an error (here, the model before planting) writes a better key for that error than the model itself, even though its implementations sometimes crash. Without knowing the errors in advance, the audit has to come from probes like ours.
\item \emph{Check generated inputs directly.} Whether a task has a requested property should be read off its inputs, not inferred from its keys. Our key-agreement proxy for anchors was wrong on every checkable task it selected.
\item \emph{Watch the pass band.} A belief held near 100\% gets almost no direct gradient under rejection sampling, whatever the key. Some exploration (more samples, hints or a teacher demonstration) is likely needed to start the fix; we did not test it.
\item \emph{Score fixes on neighbours.} Report the error distribution on items where the old belief was right, not only the belief rate on items where it was wrong, and inspect the reasoning that training selected.
\end{enumerate}

\section{Related work}\label{sec-related}

\textbf{Self-play with executors and provers.} Absolute Zero ~\cite{zhao2025} proposes and solves code-reasoning tasks, with an executor that validates tasks and checks answers; its deduction task is our output prediction. Self-Challenging Agents ~\cite{zhou2025} write tasks with verification functions and tests. STP ~\cite{dong2025} alternates conjecturing and proving in Lean. These systems guarantee that the key is consistent with the program or the proof; our results concern errors the program itself encodes. Sol-Ver ~\cite{lin2025solver} and CURE ~\cite{wang2025cure} train models to write both code and unit tests, and CodeT ~\cite{chen2022} ranks generated code by agreement with model-written tests, which assumes that errors in code and tests are not correlated.

\textbf{Generated oracles share the generator's errors.} The closest prior work is in software testing. LLM-generated test oracles tend to capture what the code does rather than what it should do ~\cite{konstantinou2024}, and when tests are written after the code, its faults are copied into them: such tests catch 14\% of faults, against 25\% for independently written tests ~\cite{konstantinou2026}. \texttt{sexec} is a training-time version of this effect, in which the copied fault becomes the training signal. In self-play against reference-free LLM judges, answers wrongly accepted by one judge also fool judges from other families ~\cite{zhou2026judges}.

\textbf{Label-free rewards and their limits.} TTRL ~\cite{zuo2025} and R-Zero ~\cite{huang2025rzero} use majority votes as pseudo-labels, and majority voting can extinguish correct minority answers ~\cite{lin2026}. He et al. ~\cite{he2026} show that intrinsic rewards sharpen the initial distribution, and Huang et al. ~\cite{huang2024sharpening} analyse self-improvement as sharpening toward the model's own high-confidence outputs. Self-rewarding RL can collapse through reward hacking ~\cite{shafayat2025}, and self-improvement is bounded by the gap between generating and verifying ~\cite{song2024}. Qwen models can improve under spurious rewards ~\cite{shao2025spurious}, a reminder that generic drift, like our runtime control's, must be separated from the effect of the key. Recursive training on generated data narrows the distribution ~\cite{shumailov2024}, as our generator's output did over rounds.

\textbf{Self-training by filtering.} STaR ~\cite{zelikman2022}, ReST\textsuperscript{EM} ~\cite{singh2023} and expert iteration ~\cite{anthony2017,polu2022} fine-tune on samples that pass a binary check. Group-relative RL ~\cite{shao2024deepseekmath} gets no signal from groups that all pass or all fail, and DAPO ~\cite{yu2025} filters them out. Learnability rewards in self-play proposers ~\cite{zhao2025,huang2025rzero} favour tasks of intermediate difficulty. All of these share our pass band, and with it the block on beliefs held near 100\%. Outcome-only filtering admits correct answers reached by wrong reasoning ~\cite{uesato2022,lightman2023}, which our rationale audit finds in the S2 training data.

\textbf{Oracles, correlated errors and shortcuts.} The test-oracle problem ~\cite{barr2015} asks where correct expected outputs come from. N-version programming ~\cite{knight1986}, differential testing ~\cite{mckeeman1998} and co-training ~\cite{blum1998} depend on independence between versions or views; pseudo-labelling ~\cite{arazo2019} and studies of correlated LLM errors ~\cite{kim2025,goel2025} show what happens without it. Shortcut learning ~\cite{geirhos2020} and underspecification ~\cite{damour2020} describe models that adopt a rule the training data do not rule out, which is how we read the replacement rules. LLMs struggle to correct their own reasoning without external feedback ~\cite{huang2023selfcorrect,stechly2024}; our results add that external feedback corrects only what it can contradict and what the filter lets through.

\textbf{Planted behaviour as a model organism.} Backdoored behaviours that persist through safety training ~\cite{hubinger2024} and false facts implanted by fine-tuning on synthetic documents ~\cite{slocum2025} use planting to study what training does to a known behaviour. We plant misreadings of task specifications and ask whether a self-improvement loop removes them.

\section{Limitations}\label{sec-limits}

\begin{itemize}
\item \textbf{One model, a small loop.} We use Qwen3-4B-Base and expert iteration rather than RL, with 3,000 task requests per round (2,178–2,705 valid at round 0) and 3–6 rounds. Generation narrows over rounds and the runtime training sets shrink by a factor of about 4–8 by round 2 (spec arms 1.4–2.7); spec rounds train on 76–466 examples. A loop this small may behave differently from production-scale self-play.
\item \textbf{Two seeds, and some single runs.} Several seed-0 findings did not replicate: the runtime control's drift, \texttt{atrust} entrenchment and the R1 replacement rule. The main spec contrasts hold in all four runs, but effect sizes vary widely, and our confidence intervals do not include run-to-run variation. The six-round results, the S1 break-out and both neighbour-topic runs are single runs.
\item \textbf{Planted beliefs are template-like.} S1 was planted with one implementation, \texttt{sorted(xs)[k]}; every S2 planting example drew the limit from the list, and the QA examples stated the belief in words. The generator kept the S2 template, so the one-sidedness of S2 corrective data is partly a planting artifact. The selection argument of Section~\ref{sec-gates} does not depend on it, but the missing anchors do. The runtime beliefs are natural, which partly answers this for the runtime half.
\item \textbf{Narrow probes.} All S2 diagnostic items, like almost all generated S2 tasks, put the limit in the list, so the truth rate measures one case of the belief. S1 has no neighbour probe.
\item \textbf{One decorrelated key.} \texttt{sclean} uses the unplanted base model, which shares the S1 misreading on 5–34\% of S1 tasks. We did not test a key writer from another model family or a weaker model.
\item \textbf{One unplanted generation round.} The comparison with the unplanted generator is a single untrained round.
\item \textbf{The remedy was not tested.} The neighbour-topic runs contained no limit-absent tasks (Section~\ref{sec-anchors}). Whether true neighbours remove the replacement rule remains open; a direct test would inject script-generated anchors or filter generated tasks on their inputs.
\item \textbf{No capability measure.} We did not measure general coding ability before and after planting or training.
\item \textbf{Not pre-registered.} A design document was written first, and the build departed from it: output prediction instead of code writing; expert iteration instead of GRPO; 8 samples and a 1–7-of-8 band instead of 16 and a [0.125, 0.625] band; 3 rounds (6 in extensions) instead of 4; two seeds instead of three; no no-loop control or multi-check key arm; no capability gate, and weak planting deliberately below the planned 0.8 planting threshold (S2 at 0.75); code-writing probes, S3, the planned hypothesis tests H4–H7 and the Holm-corrected mixed-effects analysis dropped. The lab notebook records each analysis in the order it was done, including conclusions we later revised, such as an earlier reading of the neighbour-topic runs as evidence against the remedy.
\item \textbf{Incidents.} Two runs overlapped on the GPU for about 1.5 hours, causing out-of-memory kills; the affected weak-planting arms were discarded and re-run. One unexplained hard reset of the machine left three empty files, which were removed and regenerated.
\end{itemize}

\section{Conclusion}\label{sec-conclusion}

Self-play with code execution grounds a model in what code does. It does not ground it in what a task means when the model's misreading has reached its code, and it cannot directly correct a belief for which the model almost never produces a correct sample. When it did correct a belief in our runs, it did so from cases that all lay on one side of the belief, with reasoning selected by its final answer, and the model can learn a shifted rule instead of the truth. Whether a loop can correct a given error depends on its generator, its key and its pass band. Each can be audited task by task, though in our runs early counts did not predict how fast a belief moved; what the correction installs has to be checked on neighbours.

\bibliographystyle{plainnat}
\bibliography{refs}

\clearpage
\appendix

\section*{Appendix A: Task fate at round 0}

\begin{table}[!htb]\centering\small
\begin{tabularx}{\textwidth}{@{}lllYYY@{}}
\toprule
\textbf{Run / arm} & \textbf{Topic} & \textbf{tasks} & \textbf{key = truth / belief / other} & \textbf{trained, true} & \textbf{trained, other} \\
\midrule
runtime s0, \texttt{atrust} & R1 & 60 & 22 / 14 / 24 & 17 & 17 \\
runtime s0, \texttt{atrust} & R2 & 128 & 47 / 5 / 76 & 34 & 36 \\
runtime s0, \texttt{agate} & R1 / R2 & 60 / 128 & kept 22 / 47 (rest rejected) & 17 / 34 & 0 \\
runtime s0, \texttt{exec} & R1 / R2 & 60 / 128 & all true & 33 / 92 & 0 \\
spec strong, \texttt{sexec} & S2 & 461 & $<$ count on 431 of 431 checkable & 1 & 21 \\
spec strong, \texttt{sclean} & S2 & 461 & $\le $ count on 88\%; the rest exceptions & 9 & 0 \\
spec weak, \texttt{sexec} & S2 & 461 & $<$ count on 430 of 430 checkable & 2 & 88 \\
spec weak, \texttt{sclean} & S2 & 461 & $\le $ count on 94\%; the rest exceptions & 24 & 0 \\
\bottomrule
\end{tabularx}
\caption{Fate of belief-topic tasks at round 0. Runtime counts are tasks that touch the belief. For the spec arms, ``true'' in the training columns means equal to the decorrelated key; the key audits are direct counts on calls of the form \texttt{f(list, number)}.}\label{tab-fate}
\end{table}

\section*{Appendix B: S2 trajectories}

\begin{table}[!htb]\centering\small
\begin{tabularx}{\textwidth}{@{}lYYYYYYY@{}}
\toprule
\textbf{\texttt{sclean} run} & \textbf{r0} & \textbf{r1} & \textbf{r2} & \textbf{r3} & \textbf{r4} & \textbf{r5} & \textbf{r6} \\
\midrule
strong, seed 0 & 0.03/0.07 & 0.03/0.10 & 0.16/0.16 & 0.58/0.38 & 0.66/0.34 & 0.83/0.40 & 0.88/0.46 \\
strong, seed 1 & 0.03/0.06 & 0.03/0.10 & 0.05/0.11 & 0.28/0.24 &  &  &  \\
weak, seed 0 & 0.05/0.01 & 0.05/0.04 & 0.08/0.05 & 0.23/0.24 &  &  &  \\
weak, seed 1 & 0.05/0.01 & 0.08/0.05 & 0.10/0.14 & 0.29/0.18 &  &  &  \\
neighbour topics, 1st & 0.03/0.07 & 0.03/0.08 & 0.04/0.09 & 0.14/0.16 & 0.21/0.19 &  &  \\
neighbour topics, 2nd & 0.03/0.07 & 0.02/0.07 & 0.02/0.09 & 0.10/0.17 & 0.10/0.10 & 0.28/0.19 & 0.28/0.18 \\
\bottomrule
\end{tabularx}
\caption{S2 truth on diagnostic items / overcount on control items, for every \texttt{sclean} checkpoint (the data of Figure~\ref{fig-overcount}). r0 is the planted model; the neighbour-topic runs share strong seed 0's round 0.}\label{tab-traj}
\end{table}

\section*{Appendix C: Paired contrasts at round 3}

\begin{table}[!htb]\centering\small
\begin{tabularx}{\textwidth}{@{}lYYYY@{}}
\toprule
\textbf{Contrast (100 paired items)} & \textbf{strong s0} & \textbf{strong s1} & \textbf{weak s0} & \textbf{weak s1} \\
\midrule
S2 truth, \texttt{sclean} $-$ \texttt{sexec} & +0.58 & +0.27 & +0.20 & +0.26 \\
S2 truth, \texttt{sclean} $-$ \texttt{sctrl} & +0.56 & +0.26 & +0.18 & +0.26 \\
S2 belief, \texttt{sexec} $-$ planted & $-$0.10 & $-$0.10 & +0.08 & +0.13 \\
\bottomrule
\end{tabularx}
\caption{Paired item-level bootstrap (4,000 resamples). Every interval excludes zero; the closest to zero is weak seed 0, \texttt{sexec} $-$ planted, [+0.01, +0.15]. No multiplicity correction. Runtime contrasts against \texttt{ctrl} (R1 and R2 belief, two seeds): \texttt{exec} $-$0.23 to $-$0.53 and \texttt{agate} $-$0.21 to $-$0.38, all intervals below zero; \texttt{atrust} +0.30, +0.03, $-$0.15 and $-$0.16, with the second interval including zero.}\label{tab-paired}
\end{table}

\section*{Appendix D: Reproducibility}

Code, prompts, probe sets, per-checkpoint metrics and the lab notebook are available from the authors on request. Probe metrics and task fates come from one metrics file produced by a collection script. The S2 boundary, limit-absent and neighbour-topic audits, the rationale audit and the paired tests come from separate scripts that read the generated tasks, training sets and probe outputs. The round-0 key audits are recorded in the lab notebook. A round takes about 20 minutes on one GB10 (generation of 3,000 tasks and 8 solves for each valid one, then SFT); the study used 76 trained rounds. Runs were executed one at a time, because two vLLM or training jobs at once on the GPU's unified memory caused out-of-memory kills (Section~\ref{sec-limits}).

\end{document}